\documentclass[useAMS,usenatbib]{mn2e}
\usepackage{graphicx}
\usepackage{rotating}
\usepackage{longtable}
\usepackage{lscape}

\def\mbh{$M_{\rm BH}$}

\def\Civ{C\,{\sc iv}}
\def\Mgii{Mg\,{\sc ii}}
\def\FeII{Fe\,{\sc ii}}
\def\Oiii{[O\,{\sc iii}]}
\def\Hb{H$\beta$}
\def\lsim{\mathrel{\rlap{\lower 3pt \hbox{$\sim$}} \raise 2.0pt \hbox{$<$}}}
\def\gsim{\mathrel{\rlap{\lower 3pt \hbox{$\sim$}} \raise 2.0pt \hbox{$>$}}}
\def\sline{$\sigma_{\rm line}$}

\title[On the geometry of broad emission in quasars]{On the geometry of broad emission region in quasars}

\author[Decarli et al.]{R. Decarli,$^1$\thanks{roberto.decarli@mib.infn.it} M. Labita,$^1$ A. Treves,$^1$ R. Falomo$^2$\\
             $^1$ Universit\`{a} degli Studi dell'Insubria, via Valleggio 11,
             22100 Como, Italy \\
             $^2$ INAF - Osservatorio Astronomico di Padova, Vicolo dell'Osservatorio 5,
             35122, Padova, Italy\\
       }

\pagerange{\pageref{firstpage}--\pageref{lastpage}} \pubyear{2008}

\begin{document}
\maketitle

\begin{abstract}
We study the geometry of the \Hb{} broad emission region by
comparing the \mbh{} values derived from \Hb{} through the virial
relation with those obtained from the host galaxy luminosity in a
sample of 36 low redshift ($z\sim 0.3$) quasars. This
comparison lets us infer the geometrical factor $f$ needed to
de-project the line-of-sight velocity component of the emitting gas.
The wide range of $f$ values we found, together with the strong
dependence of $f$ on the observed line width, suggests that a
disc-like model for the broad line region is preferable to an
isotropic model, both for radio loud and radio quiet quasars.
We examined similar observations of the \Civ{} line and found no
correlation in the width of the two lines. Our results
indicate that an inflated disc broad line region, in which the
Carbon line is emitted in a flat disc while \Hb{} is produced in a
geometrically thick region, can account for the observed differences
in the width and shape of the two emission lines.
\end{abstract}


\section{Introduction}

Super-massive black holes (BHs) are found in virtually all massive
spheroids (\citealt{KR95}). In the Local Universe BH mass
measurements can be performed through their imprint on the stellar
kinematics (see \citealt{Ferr06} and references therein). The masses
of the BHs are correlated with some large scale
properties of their host galaxies (\citealt{FM00}; \citealt{Gebhardt00};
\citealt{Magorrian}; \citealt{MH03}; \citealt{Graham07a}). The reader is
referred to \cite{Graham07b} for an up-to-date review on this topic,
and to \cite{King05} and references therein for interpretative models
of the scaling relations.

In Type--1 AGNs the line emission from the gas inside the BH radius
of influence is observed. If non-gravitational motions are
neglected, the cloud velocity at a given radius is fixed by the BH
mass \mbh. Emission lines are Doppler broadened according to the gas
motions. A simple, isotropic model of the broad line region (BLR) is
usually adopted (e.g., \citealt{Salviander07} and the references
therein); \cite{McLureDunlop02}, \cite{Dunlop03} and
\cite{Laor06} found that a disc model is preferable, and
\cite{NLS1} proved that inclination may explain some peculiar
characteristics of the so--called Narrow Line Type--1 AGNs. On the
other hand, some authors questioned the underlying virial assumption
(\citealt{Bottorf97}), at least for some broad emission lines (e.g.,
\citealt{BaskinLaor05}). \cite{McLure01} and \cite{Dunlop03}, and,
more recently, a number of reverberation mapping campaigns (e.g.,
\citealt{Metzroth06}; \citealt{Bentz07}; \citealt{Sergev07}) found
rough agreement between the virial estimate of the BH mass from
\Hb{} width and the one based on the host galaxy luminosities, but
the observed dispersions are significant. If the scatter is due to
the assumed gas dynamical model, constrains on the BLR geometry can
be inferred. \cite{Onken04} compared the virial estimates of
\mbh{} in few, well studied nearby AGNs with the stellar velocity
dispersion of their host galaxies. The large offset observed with
respect to the relation observed in inactive galaxies suggested that
an isotropic geometry is not successful, but they could not put a
better constrain on the gas dynamics due to the large uncertainties
and the poor statistics. In radio loud quasars (RLQs), the width of 
the broad lines is roughly correlated to the core-to-lobe power ratio
index, $R_{\rm c-l}$ (e.g.,\citealt{Wills86};
\citealt{Brotherton96}; \citealt{Vester00}). This dependence is
usually interpreted in terms of a flat BLR, given that $R_{\rm c-l}$
is related to the inclination angle of the jet axis with respect to
the line of sight. Whether a flat BLR model can be valid also for
radio quiet quasars (RQQs) is still unclear.

\begin{table*}
\begin{center}
\caption{Sample properties and imaging data from literature. (1),
(4)-(6): object name, redshift, radio loudness (`Q'=radio quiet;
`L'=radio loud) and $V$ apparent magnitude from Veron-Cetty \& Veron
(2006). (2)-(3): target coordinates from NED. (7): host galaxy absolute
$R$ magnitude corrected as described in section \ref{subsec_hostmag}.
(8): estimated \mbh{} using equation \ref{eq_bett}. } \label{tab_absolute} \scriptsize
\begin{tabular}{lccccccc}
\hline \noalign{\smallskip}
 Object     & R.A.          & Decl.     & $z$  & Radio &  $V$   & $M_R$  & log \mbh{}  \\    %
 name       & (J2000.0)     & (J2000.0) &      &       & [mag]  & [mag]  & [M$_{\odot}$] \\   %
 (1)        & (2)           & (3)       & (4)  & (5)   &  (6)   &  (7)   & (8)       \\   %
\noalign{\smallskip} \hline \noalign{\smallskip}                              %
   0054+144 &   00 57 09.9  &+14 46 10 & 0.171 &  Q    & 15.70  & -22.48 &   8.6   \\         %
  0100+0205 &   01 03 13.0  &+02 21 10 & 0.393 &  Q    & 17.51  & -22.00 &   8.4   \\
   0110+297 &   01 13 24.2  &+29 58 15 & 0.363 &  L    & 17.00  & -22.85 &   8.8   \\         %
   0133+207 &   01 36 24.4  &+20 57 27 & 0.425 &  L    & 18.10  & -22.69 &   8.7   \\         %
   3C48     &   01 37 41.3  &+33 09 35 & 0.367 &  L    & 16.20  & -24.76 &   9.8   \\         %
   0204+292 &   02 07 02.2  &+29 30 46 & 0.109 &  Q    & 16.80  & -22.80 &   8.8   \\         %
   0244+194 &   02 47 40.8  &+19 40 58 & 0.176 &  Q    & 16.66  & -22.29 &   8.5   \\         %
   0624+6907&   06 30 02.5  &+69 05 04 & 0.370 &  Q    & 14.20  & -24.53 &   9.6   \\         %
  07546+3928&   07 58 00.0  &+39 20 29 & 0.096 &  Q    & 14.36  & -24.08 &   9.4   \\         %
   US1867   &   08 53 34.2  &+43 49 02 & 0.513 &  Q    & 16.40  & -23.39 &   9.1   \\         %
   0903+169 &   09 06 31.9  &+16 46 11 & 0.412 &  L    & 18.27  & -22.76 &   8.8   \\         %
   0923+201 &   09 25 54.7  &+19 54 05 & 0.190 &  Q    & 15.80  & -22.15 &   8.4   \\         %
 0944.1+1333&   09 46 52.0  &+13 20 26 & 0.131 &  Q    & 16.05  & -23.23 &   9.0   \\         %
   0953+415 &   09 56 52.4  &+41 15 22 & 0.234 &  Q    & 15.30  & -22.24 &   8.4   \\         %
   1001+291 &   10 04 02.5  &+28 55 35 & 0.330 &  Q    & 15.50  & -23.42 &   9.1   \\         %
   1004+130 &   10 07 26.1  &+12 48 56 & 0.240 &  L    & 15.20  & -23.10 &   8.9   \\         %
   1058+110 &   11 00 47.8  &+10 46 13 & 0.423 &  L    & 17.10  & -22.46 &   8.6   \\         %
   1100+772 &   11 04 13.7  &+76 58 58 & 0.315 &  L    & 15.72  & -23.55 &   9.1   \\         %
   1150+497 &   11 53 24.4  &+49 31 09 & 0.334 &  L    & 17.10  & -23.66 &   9.2   \\         %
   1202+281 &   12 04 42.1  &+27 54 11 & 0.165 &  Q    & 15.60  & -22.40 &   8.6   \\         %
   1216+069 &   12 19 20.9  &+06 38 38 & 0.331 &  Q    & 15.65  & -22.29 &   8.5   \\         %
   Mrk0205  &   12 21 44.0  &+75 18 38 & 0.071 &  Q    & 15.24  & -22.63 &   8.7   \\         %
   1222+125 &   12 25 12.9  &+12 18 36 & 0.411 &  L    & 17.86  & -23.22 &   9.0   \\         %
   1230+097 &   12 33 25.8  &+09 31 23 & 0.415 &  Q    & 16.15  & -23.89 &   9.3   \\         %
   1307+085 &   13 09 47.0  &+08 19 49 & 0.155 &  Q    & 15.10  & -21.89 &   8.3   \\         %
   1309+355 &   13 12 17.8  &+35 15 21 & 0.184 &  L    & 15.64  & -23.29 &   9.0   \\         %
   1402+436 &   14 04 38.8  &+43 27 07 & 0.323 &  Q    & 15.62  & -22.96 &   8.8   \\         %
   1425+267 &   14 27 35.5  &+26 32 14 & 0.366 &  L    & 15.68  & -23.04 &   8.9   \\         %
   1444+407 &   14 46 45.9  &+40 35 06 & 0.267 &  Q    & 15.70  & -22.66 &   8.7   \\         %
   1512+37  &   15 14 43.0  &+36 50 50 & 0.371 &  L    & 16.27  & -23.09 &   8.9   \\         %
    3C323.1 &   15 47 43.5  &+20 52 17 & 0.266 &  L    & 16.70  & -23.06 &   8.9   \\         %
   1549+203 &   15 52 02.3  &+20 14 02 & 0.250 &  Q    & 16.40  & -21.86 &   8.3   \\         %
   1635+119 &   16 37 46.5  &+11 49 50 & 0.146 &  Q    & 16.50  & -22.40 &   8.6   \\         %
   3C351    &   17 04 41.4  &+60 44 31 & 0.372 &  L    & 15.28  & -23.55 &   9.1   \\         %
   1821+643 &   18 21 57.3  &+64 20 36 & 0.297 &  Q    & 14.10  & -24.44 &   9.6   \\         %
   2141+175 &   21 43 35.5  &+17 43 49 & 0.213 &  L    & 15.73  & -23.13 &   8.9   \\         %
   2201+315 &   22 03 15.0  &+31 45 38 & 0.295 &  L    & 15.58  & -24.28 &   9.5   \\         %
    2247+140&   22 50 25.3  &+14 19 52 & 0.235 &  L    & 16.93  & -23.11 &   8.9   \\         %
\noalign{\smallskip} \hline
\end{tabular}
\end{center}
\end{table*}

\Hb{} is the best studied emission line for low-redshift objects,
while \Mgii${}_{\lambda 2798}$ and \Civ${}_{\lambda 1549}$ are often
chosen for higher redshift (e.g. \citealt{McLureJarvis02},
\citealt{Peterson04}, \citealt{Kaspi05} and 2007, \citealt{Peng06}),
since they fall in the optical range for $z\gsim0.4$ and $z\gsim1.6$
respectively. \Mgii{} and \Hb{} widths are well correlated
(\citealt{Salviander07}), while \Civ{} line shows systematic
deviances from the \Hb{} values (\citealt{BaskinLaor05},
\citealt{Vester06}). \cite{Labita} studied \mbh{} derived from
\Civ{} width as a function of a $L_{\rm bulge}$-based \mbh{} for a
sample of low-redshift quasars. They found a significantly better
correlation than that reported by \cite{McLure01} for \Hb{} data.

In this paper we study the BLR geometry and gas dynamics by
comparing the \mbh{} values derived from \Hb{} broad emission with
those obtained from the host galaxy luminosity in a sample of
36 low redshift ($z\sim0.3$) quasars. The sample is selected in
order to provide similar numbers of RLQs and RQQs, so that
conclusions on the geometry of the BLR can be drawn for both
classes. Comparing our results with those of \cite{Labita}, based on
\Civ{} line, we sketch a picture of the dynamics of the gas around
the BH. Note that since the dynamical model of the BLR is assumed to
be independent of redshift, we also test the evolution of the
BH--host galaxy relations.

We define our sample in section \ref{sec_sample}. Data sources are
summarized in section \ref{sec_data}. Data analysis is described in
section \ref{sec_analysis}. We then discuss our results (section
\ref{sec_discussion}). The sketch of the broad line region dynamics
is presented in section \ref{sec_models}. Throughout the paper we
adopt a concordance cosmology with $H_0= 72$ km/s/Mpc, $\Omega_m =
0.3$ and $\Omega_\Lambda = 0.7$. We converted the results of other
authors to this cosmology when adopting their relations and data.

\section{The sample}\label{sec_sample}

We selected all quasars in the \cite{VCV06} catalogue that have been
imaged by {\it HST}-WFPC2 (exposure time $> 1000$ s). The host
galaxies are required to be elliptical, and therefore the bulge
component practically coincides with the whole galaxy. We considered
all the quasars with $z<0.6$, so that the
\Hb${}_{\lambda\,4861}$--\Oiii${}_{\lambda\lambda\,4959,5007}$ lines
are present in the optical spectra. For their observability from the
Northern hemisphere, we selected only objects with $\delta>0^\circ$.
The entire sample thus consists of 53 targets. 12 of them have
available spectra in the Sloan Digital Sky Survey (SDSS; see section
\ref{subsec_specdata}). Other 32 spectra were taken on purpose
at the Asiago Observatory (see section \ref{subsec_specdata}),
including 6 targets already observed by the SDSS. Two observed
targets were then excluded from our analysis: 0903+169 (RQQ) because
of the low signal to noise ratio in the available spectrum,
and 0923+201 (RLQ) because of the peculiar, composite profile of
its broad lines, possibly due to the interaction with a nearby
galaxy (see \citealt{McLeod94}; \citealt{Bennert07}). Thus
$\sim70$ per cent of the sample was covered, including 16 RLQs
and 20 RQQs. Table \ref{tab_absolute} summarizes the main
properties of the objects in our sample.

\section{Data sources}\label{sec_data}

\subsection{Host galaxy magnitudes and \mbh}\label{subsec_hostmag}

The host galaxy apparent $R$ magnitudes were taken from
the literature (\citealt{Hooper97}; \citealt{Boyce98};
\citealt{Kirhakos99}; \citealt{Pagani03}; \citealt{Dunlop03};
\citealt{Labita}) or converted from published $V$ or $F702W$
luminosities (\citealt{Bahcall97}; \citealt{Hamilton02};
\citealt{Floyd04}). Corrections for galactic extinction are from
\cite{Schlegel}. To perform colour and $k$-correction transformations
we adopted an elliptical galaxy template (Mannucci et al. 2001), assuming
that the host galaxies are dominated by old stellar population. The
$k$-correction for an elliptical galaxy at $z=0.3$ observed in the $R$-band
is $0.3$ mag. The passive evolution of the host galaxies follows Bressan,
Chiosi \& Fagotto (1994), as discussed in \cite{Labita}. Typical
corrections for the passive evolution are $\sim -0.3$ mag. The resulting
$R$-band absolute magnitudes of the host
galaxies are given in table \ref{tab_absolute}.

We use the relationship obtained by \cite{Bettoni03}, corrected for
the chosen cosmology, in order to estimate \mbh{} from the host
galaxy luminosity:
\begin{equation}\label{eq_bett}
\log {M}_{\rm BH}=-0.50 M_R-2.60
\end{equation}
where $M_R$ is the absolute magnitude of the bulge component of the
host galaxy. The rms of this fit is $0.39$. Table \ref{tab_absolute}
also lists the resulting \mbh{} values.

\subsection{Spectroscopic observations}\label{subsec_specdata}

32 optical spectra were taken with the $1.82$m
Cima Ekar telescope at the Asiago Observatory. The Asiago Faint
Object Spectrograph Camera was mounted in longslit spectroscopy
configuration with grisms n. 4, 7 and 8, yielding spectral
resolutions of $R\sim 300$, $555$ and $900$ ($2.10"$ slit) in the
spectral range $3500$--$7800$ \AA{}, $4300$--$6500$ \AA{} and
$6200$--$8050$ \AA{} respectively ($\Delta\lambda/$pxl = $4.24$,
$2.10$ and $1.78$ \AA/pxl). At $\lambda \approx 5000$ \AA{} the
spectral instrumental resolutions are $\sim 17$, $9.1$ and $5.5$
\AA, tight enough to distinguish prominent \Hb{} narrow emission
from the broad one.

The standard IRAF procedure was adopted in the data reduction. The
\verb|ccdred| package was employed to perform bias subtraction, flat
field correction, image alignment and combination. Cosmic rays were
eliminated by combining 3 or more exposures of the same objects, and
applying \verb|crreject| algorithm while averaging. When only one or
two bidimensional spectra were available, we applied
\verb|cosmicrays| task in the \verb|crutils| package. In order to
prevent the task from altering the narrow component of emission
lines, we masked the central region of our bidimensional spectra.
The spectra extraction, the background subtraction and the
calibrations both in wavelength and in flux were performed with
\verb|doslit| task in \verb|kpnoslit| package, using a Hg-Cd and Ne
lamps and spectrophotometric standard stars as reference. Wavelength
calibration residuals are around $0.17$ \AA{} (sub-pixel), thus
implying a negligible ($<1$ per cent) error on redshift estimates.
Absolute calibration of spectra was corrected through the photometry
of field stars, by comparing corollary imaging with Johnson's R and
V filters to the magnitudes published in the U.S. Naval Observatory
catalogue. The uncertainty in the flux calibration is $0.1$ mag.
Galactic extinction was accounted for according to Schlegel, et al.
(1998) , assuming $R_V = 3.1$.

Table \ref{tab_sample} summarizes the observed targets.

\begin{table}
\begin{center}
\caption{Sample objects spectroscopically observed at the Asiago
 Observatory. Redshifts and $V$ magnitudes are
 taken from Veron-Cetty \& Veron (2006). In the ``Available
 spectra'' column, A$x$=Asiago + grism n.$x$; S=SDSS; H={\it HST}-FOS
 archive data analyzed in Labita et al. (2006). Seeing is measured on
 corollary $R$-band imaging.
 These images were not available for objects 0100+020 and 3C48.}
\label{tab_sample}
\scriptsize
\begin{tabular}{lccccc}
\hline
\noalign{\smallskip}
 Object     & z     &  $V$   &Available& Date   & Seeing  \\
 name       &       &  [mag] &spectra  &        & [arcsec]\\
 (1)        & (2)   &  (3)   &   (4)   &  (5)   & (6)     \\
\noalign{\smallskip}
\hline
\noalign{\smallskip}
   0054+144 & 0.171 & 15.70 &  A4,S   & 13/09/06 & 1.4  \\
   0100+020 & 0.393 & 16.39 &  A4,H   & 18/09/07 & n/a  \\
   0110+297 & 0.363 & 17.00 &  A4     & 09/12/05 & 2.1  \\
   0133+207 & 0.425 & 18.10 &  A4,H   & 27/11/06 & 2.0  \\
       3C48 & 0.367 & 16.20 &  A4     & 11/12/05 & n/a  \\
  0204+2916 & 0.109 & 16.80 &  A4     & 15/10/06 & 1.7  \\
   0244+194 & 0.176 & 16.66 &  A4     & 16/10/06 & 1.7  \\
  0624+6907 & 0.370 & 14.20 &  A4,H   & 27/10/06 & 2.6  \\
 07546+3928 & 0.096 & 14.36 &  A4     & 24/12/05 & 1.9  \\
     US1867 & 0.513 & 16.40 &  A4,S,H & 27/11/06 & 1.3  \\
0944.1+1333 & 0.131 & 16.05 &  A4     & 25/04/06 & 2.4  \\
   0953+415 & 0.234 & 15.30 &  A4     & 07/02/06 & 3.1  \\
   1001+291 & 0.330 & 15.50 &  A4     & 19/02/06 & 2.4  \\
   1004+130 & 0.240 & 15.20 &  A4,S   & 07/03/06 & 2.2  \\
   1058+110 & 0.423 & 17.10 &  A4,S   & 26/04/06 & 2.0  \\
   1100+772 & 0.315 & 15.72 &  A4     & 08/03/06 & 2.7  \\
   1202+281 & 0.165 & 15.60 &  A4,H   & 20/02/06 & 2.3  \\
   1216+069 & 0.331 & 15.65 &  A8,H   & 12/04/07 & 1.9  \\
    Mrk0205 & 0.071 & 15.24 &  A7,H   & 11/04/07 & 1.6  \\
   1307+085 & 0.155 & 15.10 &  A7,H   & 24/04/07 & 1.6  \\
   1309+355 & 0.184 & 15.64 &  A4,H   & 20/02/06 & 2.4  \\
   1402+436 & 0.323 & 15.62 &  A4,S   & 07/03/06 & 3.6  \\
   1425+267 & 0.366 & 15.68 &  A4,H   & 20/02/06 & 2.9  \\
    3C323.1 & 0.266 & 16.70 &  A7,H   & 24/04/07 & 1.4  \\
   1549+203 & 0.250 & 16.40 &  A4     & 23/04/06 & 2.0  \\
   1635+119 & 0.146 & 16.50 &  A4     & 31/05/06 & 2.8  \\
   1821+643 & 0.297 & 14.10 &  A4,H   & 15/12/06 & 2.1  \\
   2141+175 & 0.213 & 15.73 &  A4,H   & 15/12/06 & 1.9  \\
   2201+315 & 0.295 & 15.58 &  A4,H   & 15/12/06 & 1.5  \\
   2247+140 & 0.235 & 16.93 &  A4,S,H & 12/09/06 & 1.5  \\
  \noalign{\smallskip}
\hline
\end{tabular}
\end{center}
\end{table}

The SDSS Data Release 5 (Adelman-McCarthy, et al., 2007) provides
spectra for 12 quasars in our sample. SDSS spectra have $R\sim 2000$
and a spectral range between $3800$ and $9000$ \AA. Uncertainties on
wavelength calibration amount to $0.05$ \AA, while flux calibration
formal errors account to 5 per cent. We re-observed six of these
objects (0054+144, US1867, 1004+130, 1058+110, 1402+436, 2247+140)
in order to perform a comparison. Both the specific fluxes at 5100
\AA{} and the \Hb{} broad line widths are in good agreement. Due to
their better spectral resolution, we will consider only the SDSS
spectra of these objects in our analysis.

\section{The virial determination of \mbh}\label{sec_analysis}

In the virial assumption, if the velocity $v$ of a particle orbiting
at a certain radius $R$ around the BH is known, the mass of the BH
is simply:
\begin{equation}\label{eq_virial}
M_{\rm BH}=\frac{R v^2}{G}
\end{equation}
In Type-1 AGNs expression (\ref{eq_virial}) can be evaluated at the
characteristic radius of the BLR, $R=R_{\rm BLR}$, which can be
measured almost directly with the reverberation mapping technique
(\citealt{Blandford}). \cite{Kaspi00} found that the nuclear
monochromatic luminosity, $\lambda L_{\lambda}$, is correlated with
$R_{\rm BLR}$ by $\lambda L_{\lambda}\propto R_{\rm BLR}^\gamma$,
with $\gamma$ depending on the considered emission line. The broad
line width is used to estimate $v_{\rm BLR} = v (R_{\rm BLR})$.
According to the adopted dynamical model,
\begin{equation}\label{eq_fdef}
v_{\rm BLR} = f \cdot \rm FWHM
\end{equation}
where $f$ is a geometrical factor\footnote{The reader should notice
that different definitions for $f$ are available in literature:
\cite{Vester06} and \cite{Collin} use $M_{\rm BH}=f\cdot G^{-1}
R_{\rm BLR} (\rm FWHM)^2$.} and the FWHM is expressed in velocity
units. If the gas moves isotropically, a gaussian line shape is
observed, with $f \approx \sqrt{3}/2$. If a rotational component is
present, an axial symmetry is introduced, and $f$ depends on the
inclination angle $\vartheta$:
\begin{equation}\label{eq_f2def}
f = \left(2 c_1 \sin \vartheta + \frac{2}{\sqrt{3}}c_2\right)^{-1}
\end{equation}
where $\vartheta$ is the angle between the line of sight and the
rotation axis, and $c_1$ and $c_2$ are parameters accounting for the
importance of the disc and isotropic components respectively. The
reader is referred to  \cite{McLure01}, \cite{Labita},
\cite{Collin} and \cite{NLS1} for detailed discussions on this
topic. Since $f$ value is generally unknown, one defines the Virial
Product as:
\begin{equation}\label{eq_VPdef}
VP = M_{\rm BH} \cdot f^{-2} = G^{-1} R_{BLR} (\rm FWHM)^2,
\end{equation}
corresponding to \mbh{} if $f$ is taken unitary. Thus, an estimate
of $f$ requires measures of the monochromatic luminosity $\lambda
L_\lambda$, of the line width and independently of \mbh{}.

\begin{figure}
\begin{center}
\includegraphics[width=0.49\textwidth]{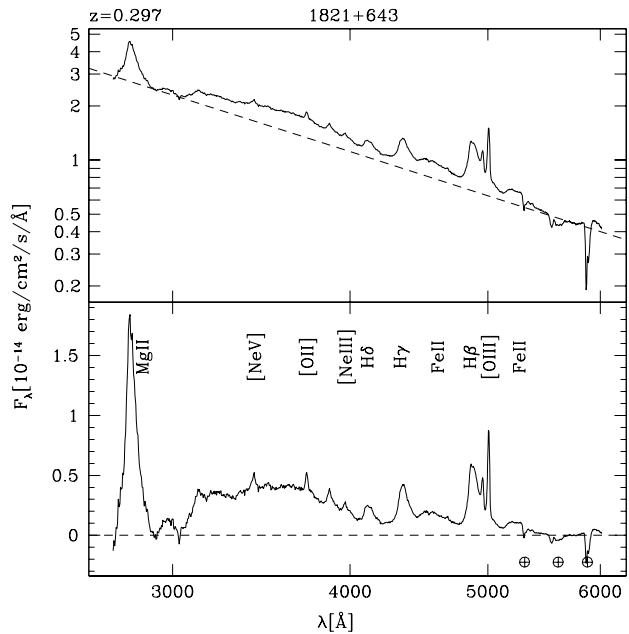}\\
\caption{The rest-frame spectrum of 1821+643. The upper panel shows
the observed spectrum with the fitted power-law for the continuum.
The lower panel shows the residual after continuum subtraction. Main
emission lines and the atmospheric absorption features are also
labeled. All the spectra are available in electronic format at
www.dfm.uninsubria.it/astro/caqos/index.html.}\label{fig_spectrum}
\end{center}
\end{figure}

\subsection{The monochromatic luminosity}\label{subsec_continua}

The continuum luminosity was calculated as follows: Rest-frame
spectral continua were fitted with a power-law (see figure
\ref{fig_spectrum}). The key-point is to avoid contaminations due to
various emission features, especially to broad \FeII{} bands.
Typical rest-frame fitted regions are around 2610, 3030, 4030, 4190,
5080 \AA{} and in the 5450-5720 \AA{} range. The $\sim$ 5100 \AA{}
region is generally free from contaminations and can be easily
fitted. The resulting fit function was computed at 5100 \AA, thus
providing $\lambda F_{\lambda}$(5100 \AA). We corrected for the
host galaxy contamination by computing the fraction of its flux
within the slit, on the basis of the nucleus and host galaxy
luminosities and the effective radius estimates available from the
literature, and following the recipe adopted by \cite{Sbarufatti}.
The average correction is $0.06$ dex.

\begin{table}
\begin{center}
\caption{Continuum luminosity parameters. (1): object name. (2):
object measured redshift. (3): data set.
`A' refers to Asiago data, the number referring to the adopted grism; `S' refers to SDSS data. (4):
$\log F_\lambda(5100 \AA)$, computed as described in section \ref{subsec_continua}.
(5): final $\log \lambda L_{\lambda}$ [erg/s]. (6): computed $\log
R_{\rm BLR}$ [cm].} \label{tab_flux} \scriptsize
\begin{tabular}{lccccc}
\hline
\noalign{\smallskip}
 Object     & z     & Data & $\log F_\lambda$ &$\log \lambda L_\lambda$ &$\log R_{\rm BLR}$\\
 name       &       & set  &[erg/s/cm$^2$/\AA]& [erg/s]            & [cm] \\
 (1)        & (2)   & (3)  &      (4)         & (5)            & (6) \\
\noalign{\smallskip}
\hline
\noalign{\smallskip}
  0054+144 &0.171   & S    & -14.9    &44.7        & 17.2 \\
 0100+0205 &0.393   & A4   & -15.6    &44.9        & 17.3 \\
  0110+297 &0.363   & A4   & -15.5    &44.7        & 17.2 \\
  0133+207 &0.424   & A4   & -15.6    &44.7        & 17.2 \\
  3C48     &0.369   & A4   & -14.9    &45.4        & 17.6 \\
  0204+292 &0.110   & A4   & -15.1    &44.2        & 16.8 \\
  0244+194 &0.174   & A4   & -15.1    &44.7        & 17.2 \\
  0624+6907&0.370   & A4   & -14.5    &46.1        & 18.1 \\
 07546+3928&0.096   & A4   & -14.2    &44.7        & 17.2 \\
  US1867   &0.515   & S    & -15.3    &45.3        & 17.6 \\
0944.1+1333&0.134   & A4   & -14.8    &44.5        & 17.1 \\
  0953+415 &0.235   & A4   & -14.8    &45.6        & 17.8 \\
  1001+291 &0.330   & A4   & -14.9    &45.1        & 17.5 \\
  1004+130 &0.241   & S    & -14.7    &45.2        & 17.5 \\
  1058+110 &0.423   & S    & -15.9    &44.6        & 17.1 \\
  1100+772 &0.311   & A4   & -14.4    &45.8        & 17.9 \\
  1150+497 &0.334   & S    & -15.6    &44.6        & 17.1 \\
  1202+281 &0.165   & A4   & -15.3    &44.6        & 17.1 \\
  1216p069 &0.331   & A8   & -14.7    &45.6        & 17.8 \\
   Mrk0205 &0.071   & A7   & -14.7    &44.5        & 17.0 \\
  1222+125 &0.412   & S    & -15.6    &44.8        & 17.2 \\
  1230+097 &0.416   & S    & -15.2    &45.2        & 17.5 \\
  1307p085 &0.155   & A7   & -15.0    &45.3        & 17.6 \\
  1309+355 &0.184   & A4   & -14.7    &44.4        & 17.0 \\
  1402+436 &0.323   & S    & -14.7    &45.4        & 17.7 \\
  1425+267 &0.366   & A4   & -15.4    &45.2        & 17.5 \\
  1444+407 &0.267   & S    & -14.9    &45.1        & 17.4 \\
  1512+37  &0.371   & S    & -15.3    &45.0        & 17.4 \\
   3C323.1 &0.266   & A7   & -15.2    &45.2        & 17.5 \\
  1549+203 &0.253   & A4   & -15.2    &44.8        & 17.3 \\
  1635+119 &0.148   & A4   & -15.4    &42.2        & 15.5 \\
  3C351    &0.372   & S    & -14.7    &45.6        & 17.8 \\
  1821+643 &0.297   & A4   & -14.2    &46.1        & 18.1 \\
  2141+175 &0.211   & A4   & -15.0    &45.0        & 17.4 \\
  2201+315 &0.295   & A4   & -14.9    &45.4        & 17.7 \\
  2247+140 &0.235   & S    & -15.3    &44.6        & 17.1 \\
\noalign{\smallskip}
\hline
\end{tabular}
\end{center}
\end{table}

\subsection{Line width measurements}\label{subsec_linewidth}

The observed \Hb{} broad component is usually contaminated by other
spectral features, in particular by blended \FeII{} multiplets, the
\Hb{} narrow component and the \Oiii$_{\lambda \lambda 4959, 5007}$
lines.

Strong \FeII{} emissions are commonly detected in quasar spectra at
4400-4750 and 5150-5450 \AA{}, and weaker blended features are found
at 4800-5000 \AA{}. A common practice to remove this
contamination (e.g., \citealt{BG92}; \citealt{Marziani};
\citealt{Salviander07}; \citealt{McGill07}) is to adopt the spectrum
of I Zw001 as a template of the \FeII{} emission, due to the
intensity and narrowness of the \FeII{} lines. Since the relative
intensities of the various \FeII{} features may differ from a quasar
to another (see, for example, table 7 in \citealt{Phillips78}), we
preferred a more conservative approach: we modeled the \FeII{}
emission as a simple power-law fitted at $\sim 4750$ and $\sim 5100$
\AA{}, and subtracted it from the observed spectrum. The reliability
of this procedure is discussed in appendix \ref{app_feii}.

Since most of our objects have the \Hb{} red wing contaminated by
[Fe\,{\sc viii}]$_{\lambda 4894}$, \FeII${}_{\lambda \lambda 4924,
5018}$ and \Oiii$_{\lambda\lambda 4959, 5007}$ lines, a reliable
study of the broad line asymmetries is extremely hard to achieve,
and is strongly dependent on the procedure adopted in removing these
contaminations. We preferred to set the same peak wavelength for
both the gaussian curves, thus neglecting line asymmetries. In
appendix \ref{app_gh} we discuss how the use of a different fitting
function, sensitive to the asymmetries in the line profile, does not
affect the estimates of the line width in a significant way.

The fit procedure was preferred to the width measurement directly
applied to the observed data (without any fit; see for example
\citealt{Collin}) since: 1) it is applicable also to low
signal-to-noise spectra; 2) it does not require an accurate modeling
of the narrow component; 3) it is reliable even in the largest
tails, where contaminations by other emission or absorption features
may be relevant.

\begin{figure}
\begin{center}
\includegraphics[width=0.49\textwidth]{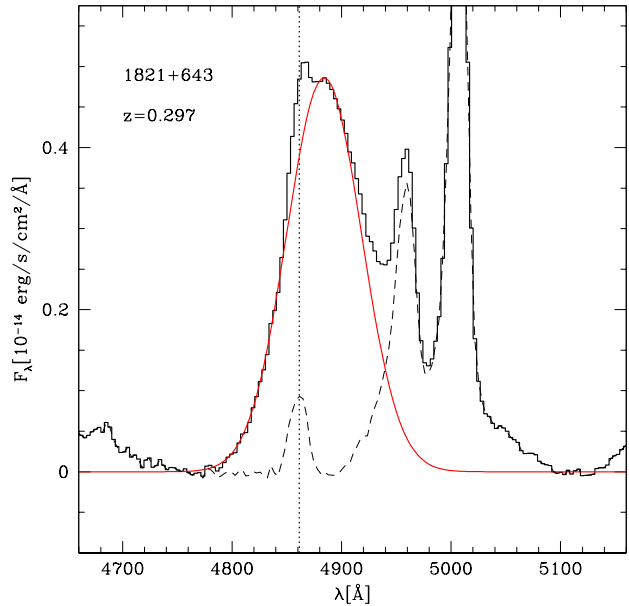}\\
\caption{The rest-frame spectrum of 1821+643 in the  4660-5160 \AA{}
range. The histogram is the observed spectrum, after continuum and
\FeII{} subtraction, as described in the text. The solid, smooth
line is the fitted curve. The residual after fit subtraction is
plotted with a dashed line. A vertical, dotted line shows the
expected \Hb{} wavelength (4861 \AA).}\label{fig_hbfit}
\end{center}
\end{figure}

We derived the FWHM and the second moment of the line, \sline, from
the fitted profile. Both FWHM and \sline{} are corrected for
instrumental spectral resolution. The ratio between FWHM and
\sline{} is used to study the shape of the line, as discussed in
section \ref{subsec_width}. According to \cite{Collin}, \sline{}
could be preferred to FWHM as a line width indicator, since it is
strongly dependent on the line wings, i.e., to the kinematics of the
innermost clouds. On the other hand, \sline{} is very sensitive to
contaminations and to the adopted \FeII{} subtraction. Therefore, we
will use the FWHM in the \mbh{} estimates, and consider \sline{}
only in the study of the line shape. Figure \ref{fig_hbfit} offers
an example of the fitting procedure, applied to 1821+643. Plots for
the whole sample are available in electronic form at
\verb|www.dfm.uninsubria.it/astro/caqos/index.html|. Our \Hb{} width
estimates are listed in table \ref{tab_check}. Typical uncertainties
in the FWHM values due to the fit procedure are $\sim10$ per cent of
the line width.

\begin{table}
\begin{center}
\caption {The estimated FWHM and \sline{} values for \Hb{}. Asiago spectra are labeled with `A$x$' in column
(2), where $x$ is the grism number. SDSS data are labeled with `S'.} \label{tab_check} \scriptsize
\begin{tabular}{lccccc}
\hline
\noalign{\smallskip}
 Object     & Data & FWHM  & \sline{} & FWHM   & \sline{} \\
 name       & set  & [\AA] & [\AA]    & [km/s] & [km/s]   \\
 (1)        & (2)  & (3)   & (4)      & (5)    & (6)      \\
\noalign{\smallskip}
\hline
\noalign{\smallskip}
   0054+144  & S  & 134&  57& 8220& 3520 \\
  0100+0205  & A4 &  79&  35& 4870& 2160 \\
   0110+297  & A4 & 100&  40& 6150& 2440 \\
   0133+207  & A4 & 143&  53& 8850& 3260 \\
       3C48  & A4 &  65&  39& 4010& 2370 \\
   0204+292  & A4 & 145&  67& 8960& 4110 \\
   0244+194  & A4 &  76&  36& 4680& 2220 \\
  0624+6907  & A4 &  59&  32& 3630& 1940 \\
 07546+3928  & A4 &  50&  28& 3110& 1750 \\
     US1867  & S  &  37&  31& 2280& 1900 \\
0944.1+1333  & A4 &  52&  30& 3230& 1850 \\
   0953+415  & A4 &  58&  47& 3610& 2890 \\
   1001+291  & A4 &  32&  26& 1980& 1610 \\
   1004+130  & S  &  98&  50& 6010& 3060 \\
   1058+110  & S  & 121&  48& 7460& 2950 \\
   1100+772  & A4 & 113&  58& 6980& 3550 \\
   1150+497  & S  &  63&  28& 3870& 1750 \\
   1202+281  & A4 &  84&  57& 5190& 3510 \\
   1216+069  & A8 &  83&  50& 5110& 3080 \\
   Mrk0205   & A7 &  53&  34& 3270& 2100 \\
   1222+125  & S  & 122&  56& 7530& 3430 \\
   1230+097  & S  &  77&  41& 4770& 2540 \\
   1307+085  & A7 &  57&  27& 3520& 1670 \\
   1309+355  & A4 &  72&  44& 4430& 2720 \\
   1402+436  & S  &  45&  33& 2760& 2030 \\
   1425+267  & A4 & 131&  73& 8090& 4530 \\
   1444+407  & S  &  44&  27& 2690& 1650 \\
    1512+37  & S  & 144&  52& 8910& 3210 \\
    3C323.1  & A7 &  77&  37& 4760& 2290 \\
   1549+203  & A4 &  30&  20& 1880& 1230 \\
   1635+119  & A4 &  92&  46& 5700& 2870 \\
      3C351  & S  & 150&  63& 9260& 3870 \\
   1821+643  & A4 &  78&  34& 4820& 2090 \\
   2141+175  & A4 &  84&  36& 5180& 2230 \\
   2201+315  & A4 &  49&  28& 3020& 1720 \\
   2247+140  & S  &  52&  22& 3180& 1330 \\
 \noalign{\smallskip}
\hline
\end{tabular}
\end{center}
\end{table}

\section{Discussion}\label{sec_discussion}

\subsection{\Hb{} line width and shape}\label{subsec_width}

The mean value and rms of the \Hb{} FWHM distribution are:
\begin{displaymath}
<\rm FWHM ~(H\beta)>_{\rm All} = 5050 \pm 2170 ~\rm km/s,
\end{displaymath}
\begin{displaymath}
<\rm FWHM ~(H\beta)>_{\rm RLQs} = 6100 \pm 2110 ~\rm km/s,
\end{displaymath}
\begin{displaymath}
<\rm FWHM ~(H\beta)>_{\rm RQQs} = 4210 \pm 1870 ~\rm km/s.
\end{displaymath}
A different distribution of FWHM is observed
between RLQs and RQQs, in the sense that the former ones show, on
average, wider lines than the latter. This difference may be
intrinsic, the velocity of the gas in RLQs being actually larger
than in RQQs, independently on the BLR geometry. Indeed, as we will
notice later, a bias towards the high--mass end of the \mbh{}
distribution occurs for RLQs (partially related to the Malmquist
bias, since the average redshift of our sample RLQs is higher than
that of RQQs). Otherwise, in the prospective of a disc--like broad
line region, different average inclination angles may account for
the different distributions in the FWHM values. This may be achieved
assuming that, while RLQs have $0^{\circ}<\vartheta<50^{\circ}$, the
RQQs are biased towards lower inclination angles, e.g.
$0^{\circ}<\vartheta<40^{\circ}$. Such a bias has been already
hypothesized by some authors (e.g., \citealt{Francis00}), but its
occurrence is still debated (see, for example,
\citealt{Kotilainen07}).

\begin{figure}
\begin{center}
\includegraphics[width=0.49\textwidth]{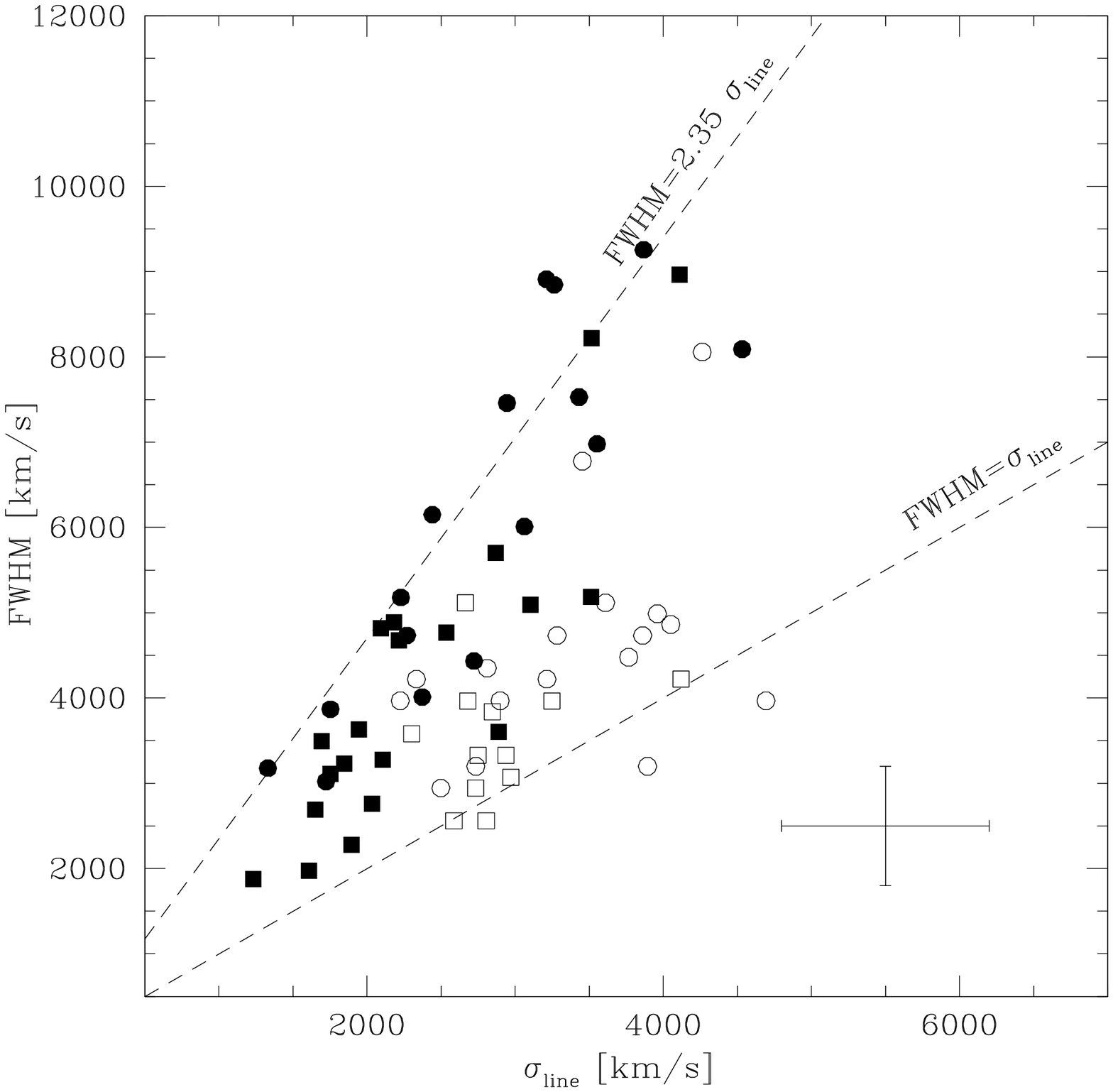}
\caption{The FWHM as a function of \sline{} for \Hb{} (filled
symbols) and \Civ{} (empty symbols). Circles refer to the RLQs,
squares to the RQQs. The dashed lines refer to FWHM$=2.35$ \sline{}
and FWHM=\sline{}. A typical error box is also plotted. \Hb{}
and \Civ{} data clearly fill different regions of the plot. No
significant difference in the FWHM/\sline{} ratio is observed as a
function of the radio loudness.}\label{fig_fwsl}
\end{center}
\end{figure}

The comparison between \Hb{} FWHM and \sline{} illustrates some
general properties of the line shape (see figure \ref{fig_fwsl}). A
single gaussian has FWHM/\sline$=\sqrt{8 \ln 2} \approx 2.35$. The
single gaussian case (upper dashed line) represents an upper limit
for FWHM. Only few \Hb{} data have FWHM$\lsim$\sline{} (lower dashed
line). \cite{Collin} suggested a bimodality in the FWHM vs \sline{}
relation, with a break when \sline{} $\gsim 2000$ km/s. We argue
that such a behaviour is mainly due to the fit procedure adopted by
those authors: When the \Hb{} width is estimated directly on the
observed spectrum, \sline{} cannot be integrated up to infinity,
because of He {\sc ii}$_{\lambda 4686}$, [Ar {\sc
iv}]$_{\lambda\lambda 4711, 4740}$ and \FeII{} contaminations.
Integral truncation leads to underestimates of the largest line
widths: For a single gaussian curve, the deviation is significant
when \sline$\gsim0.5$ times the width of the truncation interval.
Typically, \Hb{} can be studied only in the first 80 \AA{}
bluewards. That means, \sline$\gsim 2500$ \AA{} are underestimated.
In disagreement with Collin's results, our fit-based FWHM to
\sline{} ratio is found to be constant all over the observed values
of \sline. No systematic difference is reported in the
FWHM/\sline{} ratio of RLQs and RQQs.

\subsection{Broad Line Region radii}\label{subsec_radii}

Kaspi et al. (2000) found that the radius of the broad line region,
as estimated by the reverberation
mapping, is related to the continuum monochromatic luminosity
$\lambda L_\lambda$. An increasing body of measurements is now
available for \Hb{} time lags (e.g. \citealt{Kaspi00},
\citealt{Suganuma}, \citealt{Bentz06}). Following \cite{Kaspi05} we
adopt:
\begin{equation}\label{eq_kaspi05}
\frac{R_{BLR}(\rm H\beta)}{10 \rm~ light-days} = (2.00_{-0.24}^{+0.28})\cdot
\left(\frac{\lambda L_{\lambda} (5100\AA)}{10^{44} \rm~erg/s}\right)^{0.67\pm 0.07}
\end{equation}
The average characteristic radius for \Hb{} is:
\begin{displaymath}
<\log R_{\rm BLR} ~(\rm H\beta)~\rm [cm]>_{\rm All} = 17.35 \pm 0.44
\end{displaymath}
\begin{displaymath}
<\log R_{\rm BLR} ~(\rm H\beta)~\rm [cm]>_{\rm RLQs} = 17.38 \pm 0.28
\end{displaymath}
\begin{displaymath}
<\log R_{\rm BLR} ~(\rm H\beta)~\rm [cm]>_{\rm RQQs} = 17.33 \pm 0.55
\end{displaymath}
where the error is the standard deviation.

\subsection{Redshift dependence of the \mbh{}--$L_{\rm bulge}$ relation}

Our work is centered on the comparison between the Virial Products
(equation \ref{eq_VPdef}) and the \mbh{} evaluated from the host
galaxy luminosity. \cite{Woo06} and \cite{Treu07} proposed that the
\mbh{}--bulge relations change significantly even at $z\approx0.36$.
On the other hand, \cite{Lauer07} showed that such a result is
probably due to a statistical bias. Owing to the steepness of the
bright end of galaxy luminosity function, very high mass black holes
are more commonly upper residuals of the \mbh{}--host luminosity,
rather than being hosted by correspondingly very massive galaxies.
Thus, when a luminosity cutoff is adopted (typically due to
sensitivity limits, when observing at high redshift), the \mbh{}
expected from \mbh{}--$L_{\rm bulge}$ relation tends to be lower
than the real one. Since the geometrical factor is supposed to be
independent of redshift, we can directly test the evolution of the
Bettoni relation (equation \ref{eq_bett}) checking the redshift
dependence of VPs, host absolute magnitudes and their ratios. We
consider here only the objects the host galaxies of which have
similar luminosities, namely, the (15) objects with $-23>M_{\rm
R}>-24$ mag, since they are roughly well distributed along the
considered redshift range. While an overall slight increase in the
VPs is found with redshift (possibly due to Malmquist bias), data
dispersion largely exceeds the effect reported in \cite{Woo06} and
\cite{Treu07}, our objects being consistent with a no-evolution
scenario. A lower scatter is observed for \Civ{} data taken from
\cite{Labita} (see section \ref{subsec_civhb}). Applying the same
argument, no significant redshift dependence is found in the
VP-to-host galaxy luminosity ratios, the probability of null
correlation exceeding 30 per cent.

\subsection{Spectroscopic Virial Products vs Imaging $M_{\rm BH}$
estimates}\label{subsec_imaging}

\begin{table}
\begin{center}
\caption{Virial Products and geometrical factors for optical data.
(1): object name. (2): object measured redshift. (3): Virial
Products derived from \Hb{} width and 5100 \AA{} monochromatic
luminosity. Error bars are mainly due to the intrinsic scatter in
the $R_{\rm BLR}$--$\lambda L_\lambda$ relation ($\sim0.4$ dex).
(4): $f$ values.} \label{tab_VP} \scriptsize
\begin{tabular}{lccc}
\hline
\noalign{\smallskip}
 Object     & z     &log VP (H$\beta$)& $f$ \\
 name       &       &[M$_\odot$]      &     \\
 (1)        & (2)   &  (3)            & (4) \\
\noalign{\smallskip}
\hline
\noalign{\smallskip}
  0054+144   &0.171 &  8.9 & 0.7 \\
  0100+0205  &0.393 &  8.6 & 0.8 \\
  0110+297   &0.363 &  8.6 & 1.1 \\
  0133+207   &0.424 &  9.0 & 0.6 \\
  3C48       &0.368 &  8.7 & 3.0 \\
  0204+292   &0.110 &  8.6 & 1.1 \\
  0244+194   &0.174 &  8.4 & 1.1 \\
  0624+6907  &0.370 &  9.1 & 1.8 \\
 07546+3928  &0.096 &  8.0 & 4.6 \\
  US1867     &0.515 &  8.2 & 2.6 \\
0944.1+1333  &0.134 &  7.9 & 3.4 \\
  0953+415   &0.235 &  8.8 & 0.7 \\
  1001+291   &0.330 &  7.8 & 4.4 \\
  1004+130   &0.241 &  9.0 & 0.9 \\
  1058+110   &0.423 &  8.7 & 0.8 \\
  1100+772   &0.311 &  9.5 & 0.6 \\
  1150+497   &0.334 &  8.1 & 3.1 \\
  1202+281   &0.165 &  8.4 & 1.2 \\
  1216+069   &0.331 &  9.1 & 0.5 \\
  Mrk0205    &0.071 &  8.0 & 2.2 \\
  1222+125   &0.412 &  8.8 & 1.0 \\
  1230+097   &0.416 &  8.8 & 1.7 \\
  1307+085   &0.155 &  8.5 & 0.7 \\
  1309+355   &0.184 &  8.1 & 2.2 \\
  1402+436   &0.323 &  8.5 & 1.5 \\
  1425+267   &0.366 &  9.2 & 0.7 \\
  1444+407   &0.267 &  8.2 & 1.8 \\
  1512+37    &0.371 &  9.2 & 0.7 \\
  3C323.1    &0.266 &  8.8 & 1.1 \\
  1549+203   &0.253 &  7.5 & 2.4 \\
  1635+119   &0.148 &  6.9 & 0.6 \\
  3C351      &0.372 &  9.6 & 0.5 \\
  1821+643   &0.297 &  9.4 & 1.3 \\
  2141+175   &0.211 &  8.7 & 1.3 \\
  2201+315   &0.295 &  8.4 & 3.2 \\
  2247+140   &0.235 &  8.0 & 2.7 \\
\noalign{\smallskip}
\hline
\end{tabular}
\end{center}
\end{table}

\begin{figure}
\begin{center}
\includegraphics[width=0.49\textwidth]{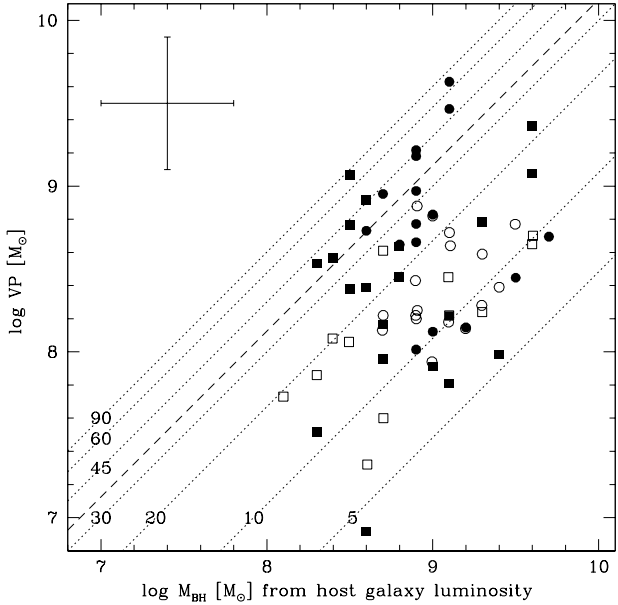}\\
\caption{Comparison between spectroscopic Virial Products based on
\Hb{} and \Civ{} and $M_{\rm BH}$ estimates based on the bulge
luminosity. Symbols are the same as in figure \ref{fig_fwsl}. The
dashed line is the expected VP as a function of \mbh{} for an
isotropic model. The values expected for a geometrically thin disc
model with different inclination angles (labeled in degrees) are
plotted as dotted lines.}\label{fig_mass}
\end{center}
\end{figure}

We are now ready to compare the Virial Products to the BH masses as
estimated from the host galaxy luminosity. VPs and $f$ values are
listed in table \ref{tab_VP}. Typical uncertainties on VPs and
\mbh{} are $\sim0.4$ dex, due mainly to the scatter in the $R_{\rm
BLR}$--$\lambda L_\lambda$ and \mbh--$L_{\rm bulge}$ relations. The
comparison is shown in figure \ref{fig_mass}. Our data show no
correlation: The probability of non-correlation is $\sim 40$ per
cent, with a Spearman's rank coefficient of $0.20$ and a residual
standard deviation of $\sim 0.50$ dex. The mean value of $f$ is $1.6
\pm1.1$. The dispersion in our data reflects the values obtained in
the same way by \cite{Dunlop03} (see their figure 13).

\begin{figure}
\begin{center}
\includegraphics[width=0.49\textwidth]{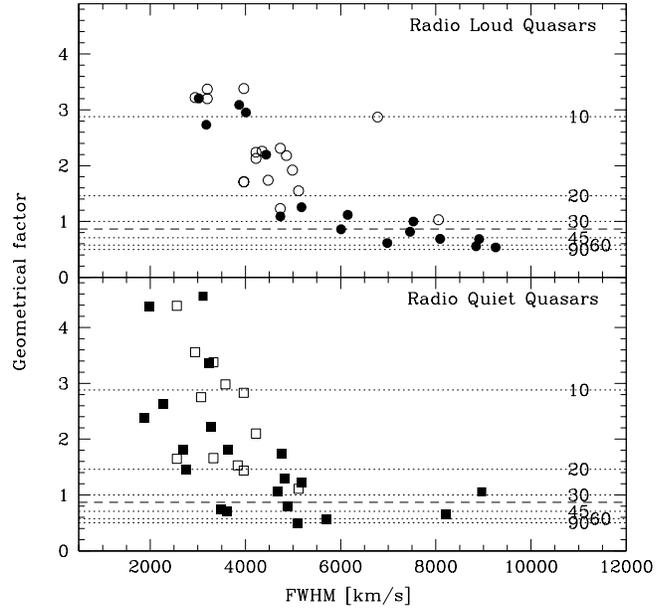}\\
\caption{Comparison between the geometrical factor, $f$, and the FWHM
for RLQs (upper panel) and RQQs (lower panel).
Symbols are the same as in figure \ref{fig_fwsl}.
The isotropic case is shown as a dashed line, while the values
of $f$ expected for a thin disc model with different inclination
angles (labeled in degrees) are plotted as dotted
lines.}\label{fig_f_FWHM}
\end{center}
\end{figure}

We study now the geometrical factor of our sample
quasars separately according to radio loudness. Indeed, indications
that RLQs may have flat BLRs have been
already reported (in particular, the rough dependence of $R_{\rm
c-l}$ on the FWHM of \Hb{}: \citealt{Wills86};
\citealt{Brotherton96}; \citealt{Vester00}), while not so much is
known about RQQs. In RLQs, the $f$
factor is found to be strongly dependent on the FWHM of the lines,
as shown in figure \ref{fig_f_FWHM}, upper panel. We note that such
a relation show significantly less dispersion than the FWHM--$R_{\rm
c-l}$ published for similar samples (e.g., see figure 4 in
\citealt{Brotherton96}). Some authors proposed the existence of two
different populations with different FWHM ranges and average $f$
(\citealt{Sulentic00}; \citealt{Collin}). Our data confirm this
result, even if a continuous trend rather than a strict bimodality
seems to be present. All the objects with FWHM$<5000$ km/s have
$f>1$, the value of $f$ rapidly decreasing with FWHM. According to
equation \ref{eq_fdef}, such a trend would be expected only if
$f$ is not fixed, given $v_{\rm BLR}$. This
reinforces the idea that RLQs have disc--like BLRs.

A similar picture is observed for RQQs, even if a larger
dispersion is found (see, for comparison, \citealt{McLureDunlop02}).
The occurrence of two populations is still clear: at FWHM$<4000$
km/s all but two targets have $f>1$, while only few targets with
FWHM$>4000$ km/s have $f>1$. This rules out that the BLR is
isotropic even in RQQs.

\begin{figure}
\begin{center}
\includegraphics[width=0.49\textwidth]{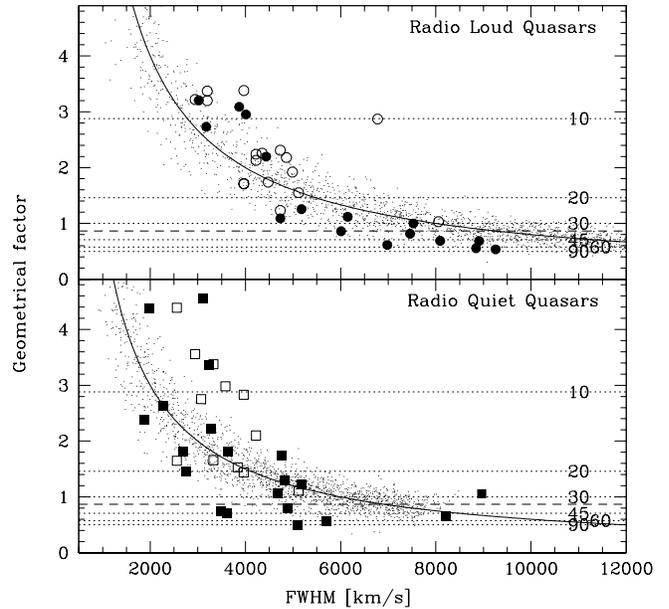}\\
\caption{The $f$--FWHM relation for RLQs (upper panel)
and RQQs (lower panel), as described in the text. 
Dots refer to the simulated values. The solid lines
show the expected $f$ assuming a thin-disc geometry with
$v_{\rm BLR}=8000$ and $6000$ km/s for RLQs and RQQs 
respectively.}\label{fig_f_FWHM_gen}
\end{center}
\end{figure}

We simulated the $f$--FWHM relation in the hypothesis of a thin disc
BLR (figure \ref{fig_f_FWHM_gen}). We assumed a gaussian
distribution of the uncertainties for FWHM, $\log$ VP and $\log
M_{\rm BH}$, mimicking the uncertainties in the adopted fitting
techniques and scaling relations. We then assumed that all the
quasars have purely disc-like BLRs, with a fixed rotational
velocity $v_{\rm BLR}$. Fitting our data with a hyperbole (see eq. 
\ref{eq_fdef}), we found $v_{\rm BLR}\sim8000$ and $6000$ km/s for 
RLQs and RQQs respectively. The variable $\vartheta$ was let free to 
vary from $0$ to $\vartheta_{\rm max}$. The angle $\vartheta_{\rm max}$ 
was fixed to $40^{\circ}$ and $50^{\circ}$ for RLQs and RQQs respectively, 
in order to match the median value of the $f$ observed distributions. 
The simulated values are overplotted to the measures presented in figure
\ref{fig_f_FWHM} for a comparison. We also plotted the two lines 
corresponding to the expected values of $f$ if uncertainties were 
negligible. Both RLQs and RQQs are well described with a disc model of the
BLR, with the former ones showing larger $v_{\rm BLR}$ than the latter 
ones. This difference cannot be explained in terms of a different range of 
$\vartheta$, and -- as we noted in section \ref{subsec_width} -- may be the 
effect of a selection bias, in the sense that the RLQs in our sample have 
higher average \mbh{} than RQQs. Concerning the different FWHM
distributions of RLQs and RQQs, our simulation cannot rule out a
dependence on the adopted range of $\vartheta$, as suggested by the
the estimates of $\vartheta_{\rm max}$ based on the distributions of $f$. 
Extending this technique on a larger sample is needed to properly address 
this topic.

\subsection{Comparison between \Civ{} and \Hb{} lines}\label{subsec_civhb}

\begin{figure}
\begin{center}
\includegraphics[width=0.49\textwidth]{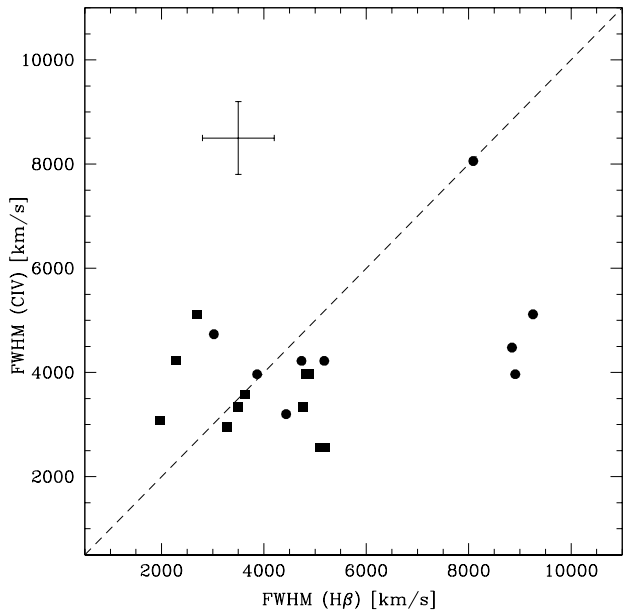}\\
\caption{Comparison between \Hb{} and \Civ{} estimates of FWHM. Circles
mark RLQs, squares refer to RQQs. A typical error box is also plotted.
The dashed line is the one-to-one relation. No
correlation is observed.}\label{fig_fw_civ_hb}
\end{center}
\end{figure}

We now want to compare the \Hb{} properties with those of the
\Civ{}$_{\lambda\,1549}$ line. We will refer to the \Civ{} data in
\cite{Labita}, who used the \Civ{} line in order to measure the VPs
of 29 low-redshift quasars observed with {\it HST}-FOS. 16 out of 29
objects in that work are also present in our sample. The average
\Civ{} FWHM and its standard deviation are:
\begin{displaymath}
<\rm FWHM ~(C{IV})>_{\rm All ~Labita ~sample} = 4200 \pm 1200 ~\rm km/s,
\end{displaymath}
\begin{displaymath}
<\rm FWHM ~(C{IV})>_{\rm Common ~sample} = 4030 \pm 1200 ~\rm km/s.
\end{displaymath}
\Civ{} FWHMs show smaller dispersion and a smaller mean value than
\Hb{} ones. This is remarkable: the \Civ{} line requires higher
ionization potential than \Hb. If the virial hypothesis is valid and
a simple photoionization model is assumed, \Civ{} emission should
show lower radii and, accordingly to equation \ref{eq_virial},
higher velocities. In section \ref{sec_models} we will show how this
point can be interpreted in terms of different thickness of the BLR
disc at \Civ{} and \Hb{} radii. Figure \ref{fig_fw_civ_hb} shows the
comparison between \Civ{} and \Hb{} FWHMs for the objects common to
both our and Labita's studies. No correlation is apparent (the
probability of non-correlation being up to $\sim 45$ per cent), as already
noticed for different samples by \cite{BaskinLaor05} (81 quasars)
and \cite{Vester06} (32 quasars; see their figure 10). We argue that
the FWHM values of the two lines are intrinsically different.

As in the \Hb{} case (figure \ref{fig_fwsl}), we now consider the
comparison between \Civ{} FWHM and \sline{}. Even if we sampled only
a small range of \sline{} values, the data from the two lines
clearly fill different regions of the (FWHM, \sline) plane. Since
the fitting procedure is similar, we can rule out a systematic
effect due to the width estimate algorithm. Thus, we argue that
\Civ{} and \Hb{} broad lines have intrinsically different shapes,
the \Civ{} departing more from the isotropic (gaussian) case than
\Hb.

The continuum monochromatic luminosity at 1350 \AA{} is also
considered for the objects belonging to the sample of \cite{Labita}.
We use this information in order to estimate $R_{\rm BLR}$ for
\Civ{} data, by means of the radius-luminosity relation
published by \cite{Kaspi07}:
\begin{equation}\label{eq_kaspi07}
\frac{R_{\rm BLR}(\rm C\,{IV})}{10 \rm~ light-days} = (0.24\pm
0.06)\cdot \left(\frac{\lambda L_{\lambda} (1350\AA)}{10^{43}
\rm~erg/s}\right)^{0.55\pm 0.04}
\end{equation}
This is the first available relation based on reverberation mapping
studies of the \Civ{} lines. However, it is based only on 8 sources,
thus the slope and offset of the relation shall need some tuning
when other data will be available. The average $R_{\rm BLR}$(\Civ{})
result:
\begin{displaymath}
<\log R_{\rm BLR} ~(\rm CIV)~\rm [cm]>_{\rm All ~Labita ~sample} = 17.19 \pm 0.31
\end{displaymath}
\begin{displaymath}
<\log R_{\rm BLR} ~(\rm CIV)~\rm [cm]>_{\rm Common ~sample} = 17.19 \pm 0.27
\end{displaymath}

The $R_{\rm BLR}$(\Civ{}) is found to be systematically ($\sim 1.7$
times) smaller than $R_{\rm BLR}$(\Hb{}), but we warn the reader
that the scatter due to the $R_{\rm BLR}$ -- $\lambda L_{\lambda}$
relations is severe. Values of $R_{\rm BLR}$(\Civ{}) smaller than
$R_{\rm BLR}$(\Hb{}) are consistent with simple photoionization
models: The \Civ{} line is a higher ionization line than \Hb, thus
it should be emitted in an inner region. Such a difference, if
confirmed, should be taken into account when referring to previous
works where the $R_{\rm BLR}$(\Hb) was used as a surrogate for
$R_{\rm BLR}$(\Civ).

Following the steps traced in section \ref{subsec_imaging}, we
estimate the VPs also for the \Civ{} data. Carbon VPs are well
correlated with imaging BH mass estimates (see figure
\ref{fig_mass}): The probability of non-correlation is $<0.1$ per
cent with a Spearman's rank coefficient of $0.63$ and a residual
standard deviation of $\sim 0.33$ dex, comparable with the
dispersion in the luminosity-radius relations. Even considering
only the common sample, the probability of non-correlation for
\Civ{} and \Hb{} data is $<0.1$ and $\sim 60$ per cent respectively.
The mean value of $f$
for \Civ{} data amounts to $2.40 \pm 0.16$. This value is $1.75$
times larger than the value obtained by \cite{Labita} on the same
data, since they adopted a different $R_{\rm BLR}$ -- $\lambda
L_\lambda$ relation (\citealt{Pian05}), which provided an estimate
of the \Hb{} broad-line radius rather than the \Civ{} one. The $f$
dependence on FWHM is observed also in \Civ{} data (see figures
\ref{fig_f_FWHM} and \ref{fig_f_FWHM_gen}). No $f$ value deduced from
\Civ{} line is found to be consistent with an isotropic model of the
BLR. The $f$--FWHM plot for \Civ{}--based data is fully consistent with
a disc-like BLR, both for RLQs and RQQs. In the average value of $f$,
the \Civ{} line appears farther from the isotropic case than the \Hb{},
both being inconsistent with isotropy.

The formally better correlation of imaging-based BH mass
estimates with VPs for \Civ{} line rather than for \Hb{} suggests
that the former could be preferable as a mass indicator than
\Hb{}, as already proposed by \cite{Vester06} and references
therein. If confirmed, this disagrees with recent claims
(\citealt{BaskinLaor05}; \citealt{Sulentic07}), in which the
average blueshift of the high-ionization lines with respect to the
rest frame of the host galaxy was interpreted in terms of gas
outflow (but see \citealt{Richards02} for a different
interpretation). A larger set of quasars with independent estimates
of both VPs and \mbh{}, sampling a wider parameter space, is
required in order to better address this topic.

\section{A sketch of the BLR dynamics}\label{sec_models}

We discuss how three simple models of the BLR dynamics can account
for or contrast with the observed \Hb{} and \Civ{} line shapes and
widths.

{\bf Isotropic model:\\}
    Up to now, since the geometry of the BLR is poorly
understood, an isotropic model has been commonly adopted as a
reference. As mentioned in section \ref{subsec_width}, if the BLR is
dominated by isotropic motions, with a Maxwellian velocity
distribution, the geometrical factor is $f=\sqrt{3}/2$ and the
FWHM/\sline{} ratio is $\sqrt{8 \ln 2}\approx 2.35$. We find that
\Hb{} FWHM/\sline{} ratios are closer to the isotropic case than
\Civ{} ones. All the geometrical factors derived from \Civ{} and
most of those from \Hb{} exceed unity, in disagreement with the
expected value for this model. Moreover, the isotropic model does
not explain the observed $f$--FWHM relation.

 {\bf Geometrically thin disc model:\\}
    If a disc model is adopted, $f$ depends on 3 free parameters,
namely $\vartheta$, $c_1$ and $c_2$ in equation \ref{eq_f2def}. Here
we assume that the disc is geometrically thin, that is, $c_2$ tends
to zero, $c_1$ tends to 1 and $f\approx (2 \sin\vartheta)^{-1}$. In
the light of the unified scheme of AGNs (\citealt{Antonucci85}),
$\vartheta$ is supposed to vary between $\vartheta_{\rm min}$, fixed
by the presence of a jet (if any) and $\vartheta_{\rm max}$, given
by the angular dimension of the obscuring torus. Even if these
angles are poorly constrained, reasonable values are
$0^{\circ}<\vartheta_{\rm min}<10^{\circ}$ and
$30^{\circ}<\vartheta_{\rm max}<55^{\circ}$ for Type-1 AGNs (see for
example the statistical approach in \citealt{Labita} and in
\citealt{NLS1}). Since the
line profile depends on the (unknown) radial distribution of the
emitting clouds, this model does not constrain the FWHM/\sline{}
ratio. The thin disc model is able to explain the $f$ values derived
from \Civ{}, with $\vartheta$ ranging between $5$ and $30$ degrees.
The $f$--FWHM relation given in figure \ref{fig_f_FWHM} is accounted
for by different inclination angles. On the other hand, \Hb{} data
are suggestive of a wider range of $\vartheta$ values, up to
$90^\circ$ in the thin disc picture. This mismatches with the Type-1
AGN model. Furthermore, objects with both UV and optical spectra
exhibit different $f$ values, for \Civ{} and \Hb{} lines. Since $f$
depends only on $\vartheta$, different inclination angles for the
orbits of the clouds emitting the two lines are needed.

{\bf Geometrically thick disc model:\\} In the case of a thick disc
model $c_2$ is non-negligible. Since \Hb{} is emitted in a larger
region than \Civ{}, the disc is thinner in the inner region (where
\Civ{} line is emitted), and thicker outside (see the {\it flared
disc} model in \citealt{Collin}, and the references therein). This
model accounts for the differences in the $f$ factor of the two
lines, and the non-correlation of the FWHMs of \Hb{} and \Civ{}
(figure \ref{fig_fw_civ_hb}). When the disc is seen almost face-on,
the velocity component perpendicular to the disc plane would be
larger than the projected rotational component, thus leading to FWHM
values larger for \Hb{} than for \Civ{}. This picture also explains
why \Hb{} FWHM/\sline{} ratios are close to the expected values for
a thermal energy distribution at a given radius from the black hole.

\section*{Acknowledgments} We thank Bradley M. Peterson, Tommaso Treu
and the anonymous referee for useful discussions and suggestions.
This work is based on observations collected at Asiago observatory.
This research has made use of the \emph{VizieR Service}, available
at \texttt{http://vizier.u-strasbg.fr/viz-bin/VizieR} and of the
NASA/IPAC Extragalactic Database (NED) which is operated by the Jet
Propulsion Laboratory, California Institute of Technology, under
contract with the National Aeronautics and Space Administration.
Observed spectra and \Hb{} fits are available at
\verb|www.dfm.uninsubria.it/astro/caqos/index.html|.

\appendix
\section{ }
\subsection{\FeII{} subtraction}\label{app_feii}

The reliability of our zero-order correction was checked as
follows:
\begin{enumerate}
\item We observed I Zw001 with the grism 7 setup. The \Hb{} line
was modeled and removed with a 2-gaussian fit.
\item For each target, we applied the zero order correction for the \FeII{}
contamination, and we fitted the broad emission of \Hb{} with a
2-gaussian profile. In order to avoid the \Hb{} narrow component, we 
excluded the line central region ($\sim1.5$ times the spectral
resolution) in the fitting procedure, when a narrow \Hb{} component
was clearly observed.
\item The spectrum of I Zw001 has been convolved to a gaussian mimicking
both the intrinsic and instrumental broadening of the \FeII{} features.
The line width was assumed to be the same as the one measured for \Hb{},
consistently with the most of the literature, thus assuming that \FeII{}
and \Hb{} emitting regions are nearly the same (but see \citealt{Kuehn08}).
\item The \FeII{} template was then scaled in flux and wavelength to match the
observed features in the $4400$--$4650$ and $5100$--$5350$ \AA{} ranges. The
best fit was subtracted to the observed spectrum.
\item The narrow lines are modeled on the \Oiii{}$_{5007 \AA{}}$ line,
following \cite{McGill07}: the \Oiii{}$_{4959 \AA{}}$ flux was assumed to 
be 1/3 of the \Oiii{}$_{5007 \AA{}}$ line. Narrow \Hb{} and He\,{\sc ii} 
fluxes are left free in the fitting procedure.
\item The resulting spectra only present the \Hb{} broad emission. We applied
a 2-gaussian fit to the \Hb{} broad component, fitting the range $4750-4975$ \AA.
\end{enumerate}
The template-subtracted FWHM and \sline{} are compared with the one
presented in the paper in figure \ref{fig_fitcheck}, upper panels.
As previously reported by \cite{McGill07}, the FWHMs obtained with
and without the subtraction of the \FeII{} template are in very good
agreement: the average difference in the two estimates is negligible
($\sim 0.03$ dex), the standard deviation being $\sim 11$ per cent
of the FWHM values, comparable to the estimated error in the fitting
procedure. The estimates of \sline{} are in agreement, too, but the
dispersion is larger: the average difference is $\sim0.04$ dex, and the
residual standard deviation is $\sim 20$ per cent.

\subsection{The fitting function}\label{app_gh}

Several authors (e.g., \citealt{McGill07}) preferred the Gauss-Hermite
series (\citealt{VanDerMarel93}) when fitting the profile of broad emission
lines. In this technique, the observed profile is fitted with a set of
orthonormal polynomials (the Hermite series) multiplied to a Gaussian curve.
\cite{VanDerMarel93} proved that, in most of the situations of astrophysical
interest, the lines are well fitted with a series truncated at the fourth
order. In this way, the set of independent parameters provided by the fit
has a straightforward interpretation, in particular $h_3$ and $h_4$ (the
coefficient for the third and fourth order of the Hermite series) are
related to the line asymmetry and kurtosis.

\begin{figure}
\begin{center}
\includegraphics[width=0.49\textwidth]{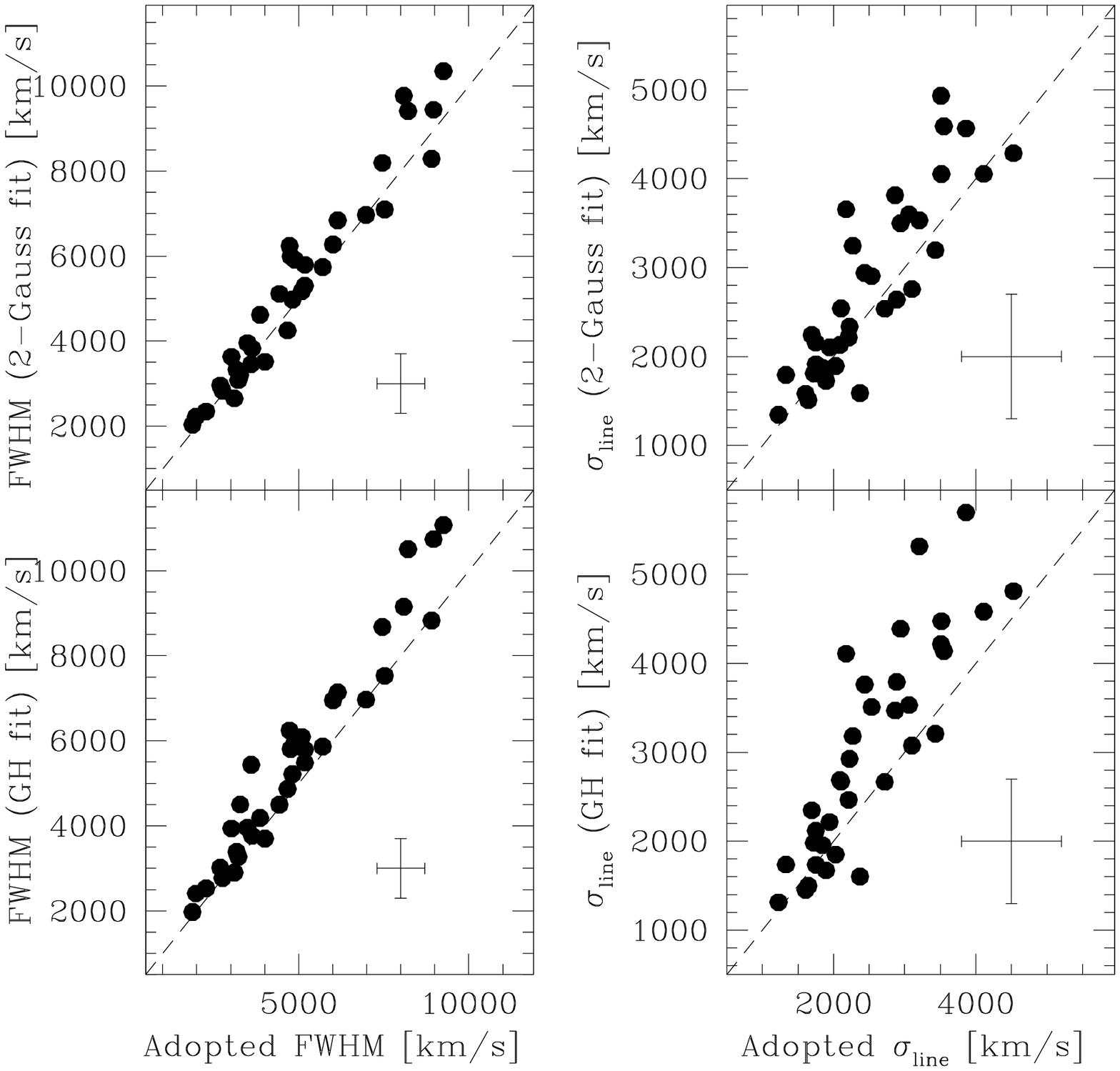}\\
\caption{The comparison between various estimates of \Hb{} FWHM
(left) and \sline{} (right). In the x-axis, the width estimates
obtained with our zero-order correction for \FeII{} emission and the
2-gaussian fit for the broad emission of \Hb{}. In the y-axis, in
the upper panel the width estimates obtained with the 2-gaussian
technique, after \FeII{} template subtraction. In the lower panel,
the width estimates based on the Gauss-Hermite fit procedure. The
dashed lines refer to the 1-to-1 case. Typical error boxes are also
shown. All FWHM estimates are consistent. As expected, the \sline{}
values are more sensitive to the correction for \FeII{} emission and
to the fitted function. }\label{fig_fitcheck}
\end{center}
\end{figure}

In order to check the dependence of our results on the adopted fitting
function, we applied the Gauss-Hermite fit (extended to the fourth order of
the series) to our data, after the \FeII{} template subtraction described in
appendix \ref{app_feii}. The FWHM and \sline{} values obtained with the two
profiles are compared in figure \ref{fig_fitcheck}, lower panels. The FWHM
estimates are in good agreement with those adopted in the paper: the average
difference ($\sim 0.05$ dex) is negligible within the purposes of our work,
with a residual standard deviation of $\sim13$ per cent. A systematic
deviation of \sline{} estimates is observed when comparing the two fitting
technique, but the overall average difference ($\sim 0.08$ dex) cannot
account for the differences in the FWHM/\sline{} ratio observed for
\Civ{} and \Hb{}.

\end{document}